\documentstyle[aps,preprint]{revtex}
\input{psfig.sty}
\begin{document}
\title{Magnetization Jump in a Model for Flux Lattice
Melting at Low Magnetic Fields}
\author{Seungoh Ryu \\ D. Stroud}
\address{Department of Physics, Ohio State University, Columbus, OH 43210}
\date{\today}
\maketitle
\begin{abstract}
Using a frustrated XY model on a lattice with open boundary conditions, we
numerically study the magnetization change near a flux lattice
melting transition at low fields.  In both two and three dimensions, we find
that the melting transition is followed at a higher temperature by the
onset of large dissipation associated with the zero-field XY transition. 
It is characterized by the proliferation of vortex-antivortex pairs (in 2D)
or vortex loops (in 3D).  At the upper transition, there is a sharp increase
in magnetization, in qualitative agreement with recent local Hall probe 
experiments.
\end{abstract}
\newpage
Among the most debated aspects of high-T$_c$ superconductors
is the nature of the superconducting transition in a magnetic field.  It is
widely believed that in the presence of disorder, the transition may fall into
the universality class of a vortex glass\cite{fisher89}, or a
Bose glass\cite{nelson92}.  In addition, disorder may produce an entirely
new transition characterized by the disappearance of the Bragg peaks of
the Abrikosov lattice above a certain magnetic field\cite{ryu96}.  
But even in the absence of disorder, the phase transition is 
not fully understood.
At high applied fields ${\bf H}$ parallel to $c$, several numerical studies 
strongly suggest that there is a {\em first-order} phase transition with
a finite latent heat\cite{hetzel92,sasik95,ryustroud96}, and 
a corresponding magnetization discontinuity\cite{sasik95}.   
At lower magnetic fields, the
evidence is more ambiguous.  At such fields, the
thermodynamics of the field-induced vortices should be strongly influenced by 
fluctuations of the zero-field $XY$ degrees of freedom, which are
far more numerous than the {\em field induced} vortex 
degrees of freedom\cite{tesanovic95}.

Experimental evidence for a first-order transition
with a magnetization jump has been reported for both 
Bi$_2$Sr$_2$CaCu$_2$O$_{8+\delta}$\cite{zeldov95} and 
YBa$_2$Cu$_3$O$_{7-\delta}$\cite{welp96}.
These measurements
show that the transition is accompanied by an {\em increase} in vortex number
density, suggesting a melting analogous to that of water.

In this paper, we present model calculations
of the superconducting-normal (SN) transition in both two and
three dimensions (2D and 3D) in the absence of disorder.  
Our results strongly suggest that in both cases, 
the SN transition {\em at low B} occurs in {\em two distinct stages}.
The first is the melting of the vortex lattice at $T_m$, 
and is characterized by disappearance of the 
Bragg peaks associated with solid
order.  The second stage is driven by the XY degrees of freedom, 
and occurs at temperatures above $T_m$.  
That zero-field XY transition is accompanied by the unbinding of
vortex-antivortex pairs (in 2D), or 
proliferation of vortex loops (in 3D)\cite{feynman55}.
In either case, these excitations screen the
repulsive interactions between the field-induced vortices or vortex
lines\cite{doniach79}. Consequently, 
the density of field-induced vortices becomes larger,
leading to an increase in the absolute magnetization as observed
experimentally.  
In 3D, this screening is
accompanied by the vanishing of $\gamma_{zz}$, the z-component of the
helicity modulus, which measures the rigidity of the phase degrees of
freedom in the $c$ direction\cite{ebner83}.  Since this occurs at a sharp 
temperature, the second stage could be a true phase transition in 3D.  

Our calculations are based on the widely-studied frustrated $XY$ model on a
stacked triangular lattice\cite{hetzel92}.  This
model has the Hamiltonian
${\cal H} = -\sum_{\langle ij \rangle}J\cos(\theta_i-\theta_j-A_{ij}),
$
where $\theta_i$ is the phase of the superconducting order parameter at
site $i$, $J$ is the isotropic coupling energy between nearest-neighbor sites,
and $A_{ij} = (2\pi/\phi_0)\int_i^j{\bf A}\cdot {\bf dl}$ 
accounts for any applied magnetic field ${\bf B}$ through a
vector potential ${\bf A}$.  
The choice of triangular
grid minimizes the periodic vortex pinning potential which
is an artifact of any discrete lattice\cite{franz94}.  
We evaluate the equilibrium and dynamical 
properties of ${\cal H}$ using respectively a standard Monte Carlo (MC)
procedure\cite{li93}, and solving the 
coupled Josephson equations with Langevin 
dynamics\cite{klee91}.  
For each plaquette $p$, we determine a vorticity
vector $n_\alpha(p)$ ($\alpha = x, y, z$) from 
$\sum_p \rm{mod} [ \theta_i - \theta_j - A_{ij}, 2\pi ]$  $= 
2 \pi [ n_\alpha(p)  - f_p ].$
Here the summation runs along the bonds $\{i,j\}$ 
belonging to the plaquette labeled
$p$; and $f_p 
\equiv \sum_p A_{ij}/(2\pi).$ 

Previous calculations\cite{hetzel92} have established that there is a {\em
first-order} phase transition at high fields in the
$c$ direction ($f \equiv \phi/\phi_0 = 1/6$, where $\phi$ is the flux per 
triangular plaquette, and $\phi_0=hc/2e$).  At low
fields (f = 1/25) on a simple cubic lattice\cite{li93},
there is an apparent double transition: 
the first stage is a melting transition, followed at
higher temperatures by a loss of phase coherence parallel to the $c$
axis (as measured by $\gamma_{zz})$. By contrast, 
Jagla {\it et al} found that for $f=1/6$, the transitions
are separate only in the presence of quenched disorder\cite{jagla96}.

\begin{figure}[b]
\begin{center}\psfig{figure=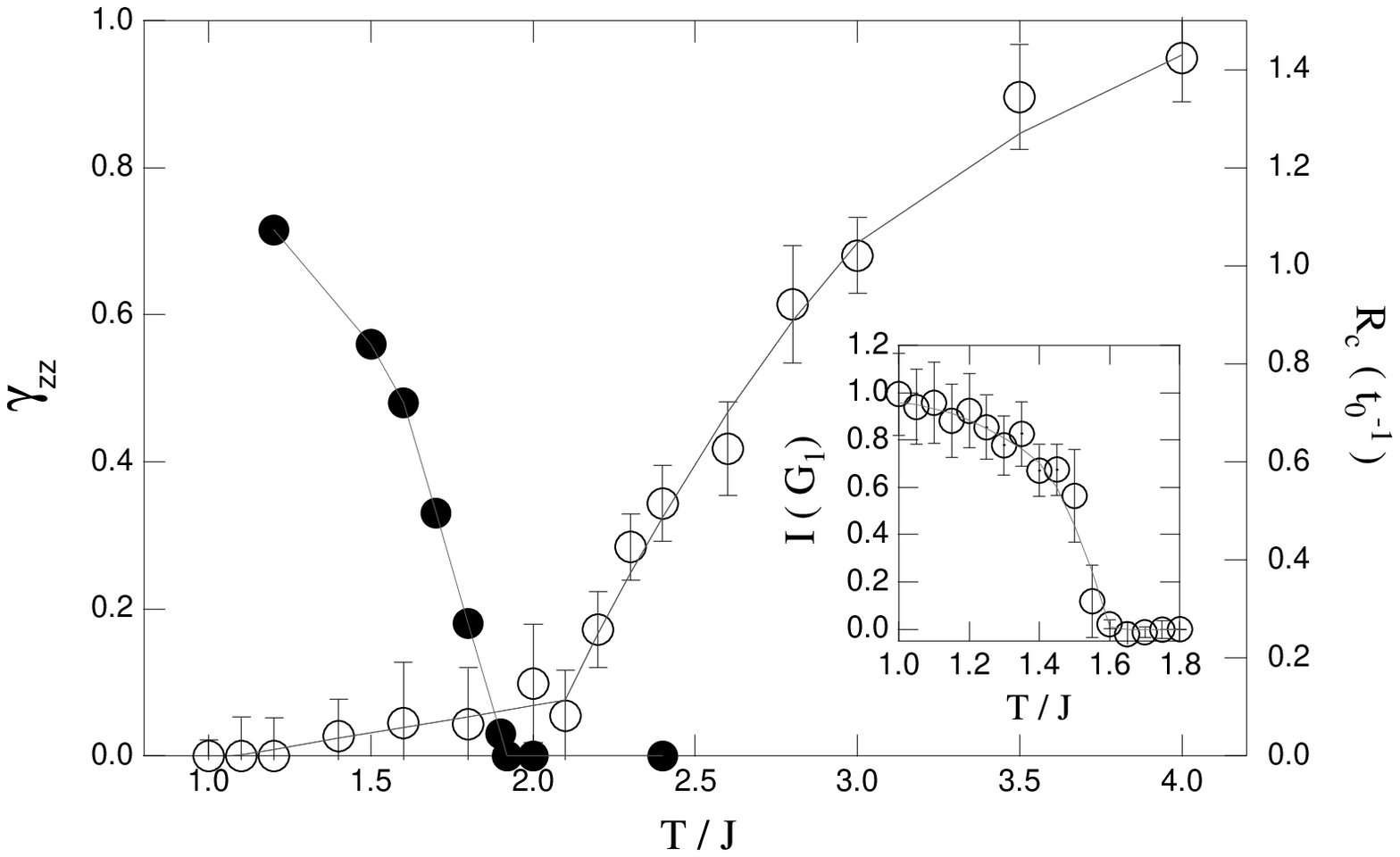}\end{center}
\caption[lala]{Helicity modulus $\gamma_{zz}$ (Filled circles); c-axis resistance 
$R_c$ (open circles) in units of $1/t_0$ where $t_0 = \hbar / (2eJR).$
The error bars for $R_c$ show the standard deviation when the averaging 
interval was varied from 
$200 t_0 $ to $600 t_0$ in steps of $10 t_0$. The inset shows the 
integrated first-order Bragg intensity I($G_1$) of the
field induced vortices, for $f=1/24.$ Error bars represent rms deviation 
over 7 different configurations 7000 MC steps apart. 
Lines are guides for the eyes.
\label{figf24}}
\end{figure}

We first show that a similar double transition occurs in the stacked
triangular lattice. 
Fig.~\ref{figf24} 
shows $\gamma_{zz}(T)$, the {\em c-axis} 
resistance $R_c(T)$, and the first order Bragg peak 
intensity $I({\bf G_1}, T)$ (inset) as a function of temperature $T$
for $f = 1/24$.  
${\bf G_1}$ is taken as one of the
six equivalent shortest reciprocal lattice vectors in the $ab$ plane.  We
use periodic boundary conditions
in all three directions and a lattice which 
is a parallelopiped of size $24 \times 24 \times 24$, i.\ e.,
48 vortex lines.  Clearly,
$I({\bf G}_1)$ and $\gamma_{zz}$
go to zero at {\em quite different} temperatures,
suggesting a double transition.  The lower temperature, denoted 
$T_m \sim 1.55 J$, corresponds to the loss of solid-like structural order.  
The upper transition occurs at $T_\ell \sim 2.0 J$. 
The behavior $T_m < T < T_\ell$ is characterized by logarithmically
slow relaxation of $\gamma_{zz}$ to its non-zero equilibrium values 
over $(5 -10)\times 10^5$ MC sweeps.
The small high-$T$ tail of $\gamma_{zz}$ depends on sample thickness,
and the data shown in the figure represent 
asymptotic behavior up to $N_z = 48$.  
Starting from typical equilibrium configurations thus obtained, we 
carried out dynamical calculations 
within the resistively shunted Josephson junction (RSJJ) model\cite{klee91,jagla96}. 
Using periodic boundary conditions, we injected a uniform current
($0.083 I_c $ per grain) through the lowermost xy-plane and extracted 
through the xy-plane in the middle of the sample, 
effectively creating a periodic pattern of alternating 
current flow along the z-axis. The resulting dissipation $R_c$ is sensitive 
mainly to the presence of {\em net transverse vorticity}, and its fluctuations.
We also calculated $R_{ab}$, analogous to the inplane resistivity, again using 
a similar periodic arrangement. 
This exerts a {\em shearing} force on two halves of the lattice and probe 
directly the shear rigidity of the lattice on length scales of our simulation 
box. 
The qualitative behavior of $R_{ab}$ is very similar to $R_{c}$, i.~e.~it 
displays a sharp increase at $T_\ell$\cite{ryu97}.

\begin{figure}[b]
\begin{center}\psfig{figure=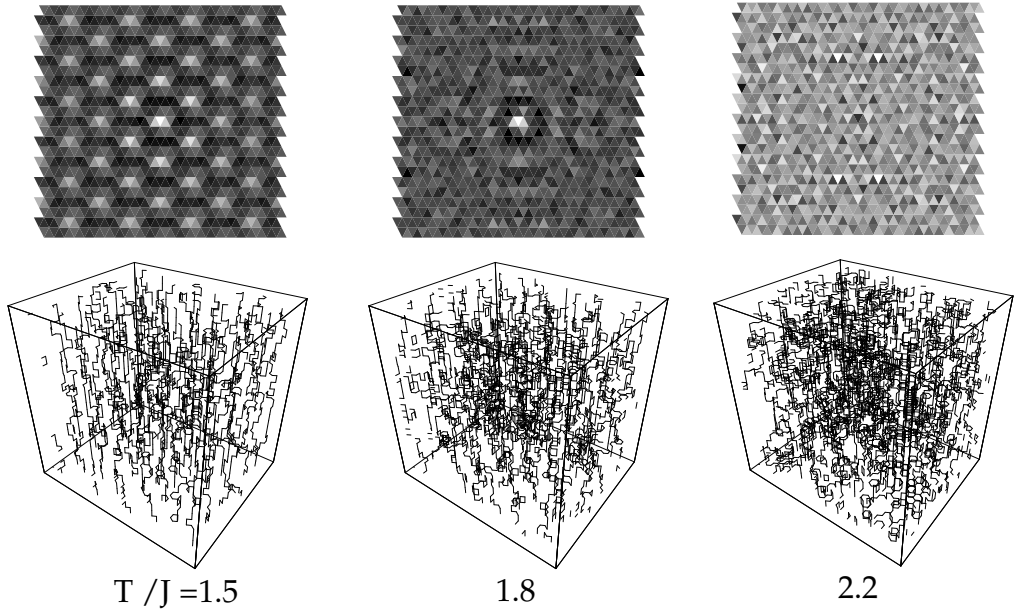}\end{center}
\caption{{\em Top-to-bottom} vortex density-density correlation function 
g$_{tb}(x,y)$ for $T/J = 1.5, 1.8$ and $2.2$ in a $24\times 24
\times 24$ system with $f=1/24.$   $T_m \sim 1.55 J$ and $T_\ell \sim 
2.0 J$. The lower panel shows typical vortex configurations.  
\label{figflux}}
\end{figure}

Fig.~\ref{figflux} 
shows the top-to-bottom vorticity density-density  correlation 
function $g_{tb}(x, y) 
\sim < n_z(x,y,z=N_z/2) n_z(0,0,0) >$ 
as a function of $T$ for $f = 1/24$.  
Note the periodic lattice of
peaks at $T/J=1.5$, the presence of only a strong central peak at $T/J = 1.8$,
and the absence of peaks at $T/J=2.2$.  The strong central peak suggests
that at $T/J = 1.8$ the vortex lattice has melted into a {\em line liquid} 
with a {\em finite} line correlation length along the field direction.
This separation between the melting and $\gamma_{zz}$
transition is a {\em low-field} phenomenon: at $f = 1/6$, the 
I(${\bf G}_1$) and $\gamma_{zz}$ 
vanish at {\em nearly the same} $T$ within our resolution.  

To probe the vortex density change associated with the transitions, 
we must abandon
periodic boundary conditions, because these enforce a fixed density 
in the system, whose value is dictated by the applied field.  
We therefore adopt {\em free transverse} boundary conditions
\cite{boundarycondition},
while still maintaining periodic boundary conditions in the $z$ direction.
This procedure allows the outermost surface of the
vortex ensemble to have a variable position.

We first discuss our results in 2D, considering $f = 1/24$ on
a triangular grid of different sizes with $N=26,52,78$ and $100$.  
One might think of taking the 
``magnetization'' $M_z$ as the average net vortex
density $n\equiv \int n_z({\bf r}) d{\bf r} / A$, 
where $n_z({\bf r})$ is the local vortex density and $A$ is the total area.  
However, $M_z$ defined in this way suffers from 
spurious boundary effects, arising from
the depletion of vortices near the boundaries in the lattice phase.   
This surface effect\cite{ebner} 
vanishes for large samples as $1 / \sqrt{A}$.
Instead, we look at the {\em local Voronoi cell area} ${\cal A}_i$, i.\ e.,
the area of the generalized Wigner-Seitz cell for
vortex $i$\cite{ryustroud96}.
A local vortex density at a point ${\bf R}$ 
in the simulation box 
may then be defined as $n({\bf R}) = \sum_i \delta( {\bf R} \in {\cal A}_i )
 / {\cal A}_i$ where $\delta ({\bf R}  \in {\cal A}_i) $ gives 1 if the 
point ${\bf R}$ lies in the Voronoi cell associated with vortex $i$. 
Next, the local magnetization $M_z$ which we equate to 
the {\em bulk average density} $<n>$ is calculated from 
$<n> = \sum_{{\bf r}_i \in {\cal C}}  { 1 \over {\cal A}_i } $
as the average of inverse Voronoi area for vortices lying within a 
measurement area ${\cal C}$ suitably distant from the sample boundary. 

For $T \ge 0.5 J,$ there are, besides the field-induced vortices,
numerous thermally induced vortex-antivortex 
dipole pairs, which must be eliminated before this
procedure is applied.  To do this, we pair each antivortex with the
nearest vortex, and remove them from the count.  Since most such dipole 
pairs are much smaller than $1/\sqrt{<n>}$, this criterion is justified.

\begin{figure}[b]
\begin{center}\psfig{figure=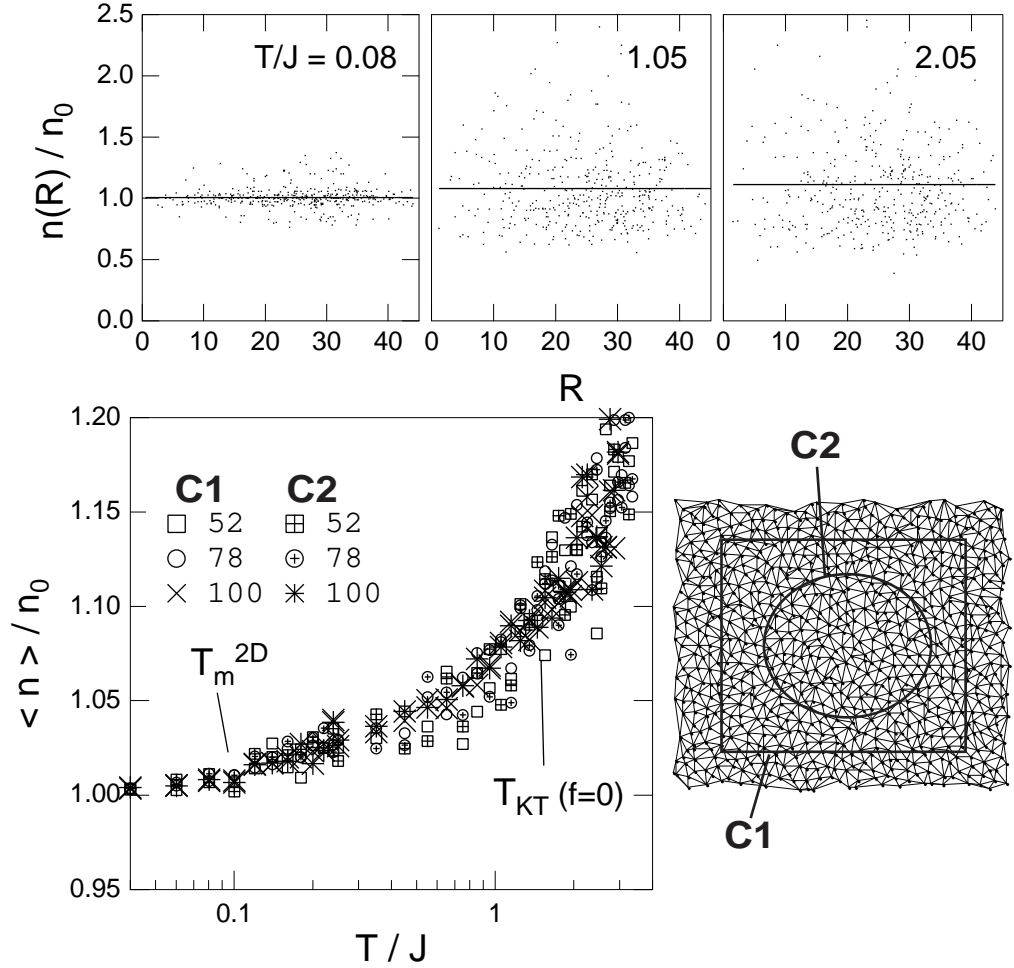}\end{center}
\caption{Upper panel:local density distribution $n({\bf R})$, plotted 
as a function of distance $R$ from the center of the plane for $T/J = 0.08, 
1.05$, and $2.05$.
Lower panel: normalized bulk vortex density vs.~T for three different sample
sizes at $f=1/24.$   Right panel: results of Delaunay triangulation for 
{\em field induced} vortices at $T/J = 1.05$.  For results
in left panel, averages were taken over vortices within 
two different bounded areas, ${\cal C}_1$ and ${\cal C}_2$.
\label{fig2d}}
\end{figure}

For $f = 1/24$ the 2D melting transition occurs
near $T_m^{2D} = 0.09 J$, as determined by the vanishing of the 
Bragg peak $I({\bf G}_1)$.  The zero-field $XY$ (i.\ e., in 2D, the
Kosterlitz-Thouless) transition occurs at about $T^0_{KT} = 1.6 J$.  
Superconductivity in 2D is destroyed at $T_m^{2D}$.
As shown in 
Fig.~\ref{fig2d}, $<n>/n_0$ (where $n_0 \equiv \sqrt{3} f / 4$ 
is the nominal density) displays
a sizeable increase across $T_{KT}^0$.   
This increase is independent of 
system size(checked up to $100 \times 100$ grains). 
As further evidence that the boundaries
are not producing this increase, we note that we obtain the
same results for $<n>$ for different choices of measurement areas, 
${\cal C}_1$ and ${\cal C}_2$.
 
The upper panel of Fig.\ 3 shows that the local density fluctuates
progressively more as $T$ increases: near $T_m$, the
rms width of the local bond length fluctuation is about 0.15,
a value close to the Lindemann number. 
The increase in density at high temperatures 
arises mainly from an increase in the frequency of short bonds,
which occur {\em throughout the sample} as is clear from the panel.
We have also observed at $T_m$ a sharp jump in the density of {\em topological 
lattice defects}, such as disclinations.  If these defects are 
predominantly of a particular type such as lattice vacancies, we would expect 
a corresponding jump in vortex density. However, we do not 
observe a sizeable increase 
in $M_z$ at $T_m$ within our resolution of about 2 \%.

As the vortex liquid is further
heated, the vortex density {\em increases} to a value well
above that of the solid. This increase is due to 
the excitation of vortex-antivortex pairs characteristic of the
$f = 0$ Kosterlitz-Thouless transition.  The excited pairs screen
the repulsion between the field-induced vortices, 
lowering the effective chemical 
potential for field induced vortices and making the system
more compressible.

\begin{figure}[t]
\begin{center}\psfig{figure=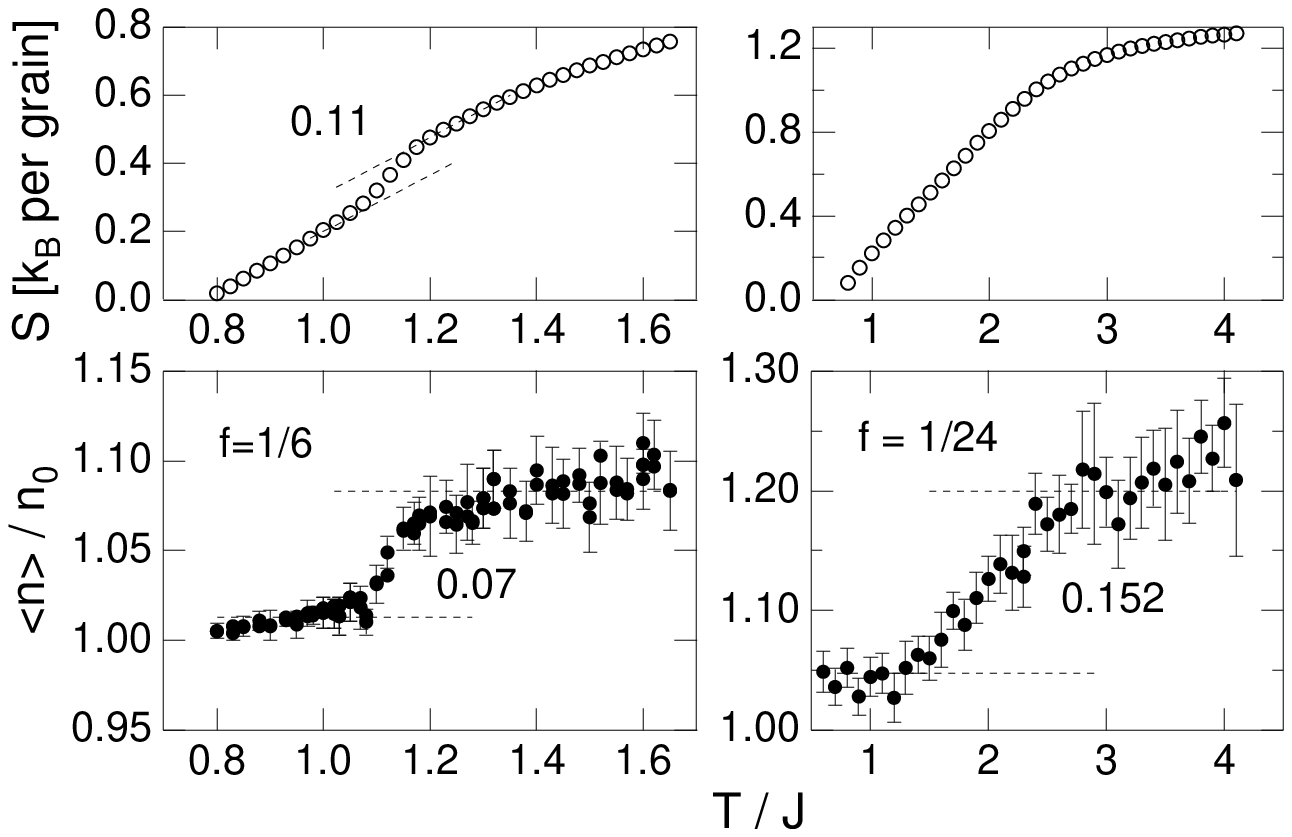}\end{center}
\caption[lalal]{Entropy $S$ (upper panel)
and normalized vortex density (lower panel) for $f=1/24$ and $1/6$ in 
$24 \times 24 \times 12$ system with open boundary conditions.
The error bars denote the rms deviations from layer to layer.}
\label{figjumps}
\end{figure}

Fig.~\ref{figjumps} shows the corresponding behavior in 3D, at $f = 1/24$
and $1/6$.  
Once again $n/n_0 \sim 1$ below the zero field 3D XY transition at $T_{XY}$, 
increasing markedly
near $T/J \sim 2.0$ for $f=1/24$ and near $1.15J$ for $f=1/6$. 
For $f=1/6$, this increases occurs over a temperature range
$\triangle T < 0.05 J$, just at the first order phase transition where 
both lattice order and phase coherence
parallel to the field are destroyed
($T_m \sim T_{\ell}$ for $f=1/6$).
The relative change in vortex density, $\delta n / n_0$, is less 
than 7 \% for $f = 1/6$.  The entropy as calculated from 
$\int^T C / T dT$ displays a corresponding jump
$\triangle S \sim 0.11 k_B$ per grain (or $0.3 k_B$ per vortex) as shown 
in the upper panel. 

For $f = 1/24$, the change in $n/n_0$ is spread over a wider temperature range 
($\triangle T \sim 0.7 J$), and the midpoint of the increase in 
$n/n_0$ falls at $T_\ell$. The overall field change $\delta B / B \sim 15 \%.$
The entropy does not show a recognizable jump near $T_m.$
For this open-boundary system, $T_m < 1.35 J$, 
about $10 \%$ below the value $T_m \sim 1.55 J$ found 
with periodic boundary conditions. 
At $f = 1/6$, the two transitions are indistinguishably 
close, and any possible change associated with melting is
overshadowed by the larger {\em increase} in $n$ due to 
screening by vortex loops at $T_\ell$ for this dense system.
Thus, for $f=1/24,$ there is a phase between $T_m$ and $T_\ell$ with
no lattice order, but yet with a diamagnetic moment almost 
as large as in the vortex lattice phase.  At high temperatures, 
the vortex density rises substantially {\em above} the solid phase.
  
This overall increase in $M_z$ is consistent with experiments, 
but the rise is associated, not 
with melting itself, but rather with the disappearance of phase coherence in 
the z direction ($\gamma_{zz}\rightarrow 0$), which is also accompanied by
dramatic changes in the transport coefficients, $R_{c}$ and $R_{ab}.$ 
In 3D, this increase is associated with the proliferation of vortex loops, 
caused by the 3D XY fluctuations.  
The loops screen the repulsion between
vortex lines, allowing them to come closer together and the net vortex
density to increase.  
In 3D, the screening also encourages collision and fusion 
of field induced vortex lines\cite{lala}, 
causing them to lose the longitudinal phase
coherence at $T_\ell$.   

The rise in $M_z$ for $f=1/24$ in our model would actually
occur over a very narrow temperature range in the real superconductor. 
This is because $J$ in YBa$_2$Cu$_3$O$_{7-\delta}$ is strongly
temperature-dependent near $T_c$ with $J \propto 
{d\phi_0^2 \over 16 \pi^3 \lambda(T)^2}$. 
Using the simple mean field form for the temperature dependence of 
$\lambda(T)$ and 
material parameters for YBa$_2$Cu$_3$O$_{7-\delta}$, we estimate 
$\triangle T \equiv [T_\ell-T_m] < 150 {\rm mK}$ for 
$f=1/24,$ and $\triangle T < 10 {\rm mK}$ for $f = 1/6$.
Thus the rise in M$_z$, in fact,
could occur over a very narrow temperature
range in high-T$_c$ materials,
in good agreement with experiments\cite{zeldov95,welp96}. 

Finally, we consider whether the upper transition in 3D might be a true,
second phase transition, and not just a crossover. Numerically, there are 
several results suggesting a separate phase transition at $T_\ell.$
First, $\gamma_{zz} \rightarrow 0$ at $T_{\ell} > T_m$\cite{li93,ryu97,koshelev97}, 
and dissipation 
due to extensive flux line cutting sets in both parallel and perpendicular 
to the field at $T_\ell > T_m$\cite{ryu97}.
An analysis of vortex configurations also reveals 
that {\em infinite} clusters of colliding vortex 
lines and loops first appear at $T_\ell$\cite{ryu97} reminiscent 
of a percolation transition.  Recent numerical simulations using 
London vortex loop model suggest a similar 
phase transition, separate from the melting transition\cite{nguyen96,chen97}.

Te\v{s}anovi\'{c} has considered the low field regime of the 
mixed state\cite{tesanovic95}. 
By a singular gauge transformation introducing fictitious 
antivortices, the original problem is transformed
into one in which the new order parameter experiences
no applied field but a fluctuating gauge field. 
The XY-type transition of this new order parameter(the 
so-called ``$\Phi$ transition'')
is a possible candidate for the transition at $T_\ell$ for $f \le 1/24$. 
We emphasize that increase in $M_z$ at T$_\ell$ results from
the onset of screening by the background vortex loop plasma.  
It appears that a similar transition between a superfluid and metallic 
state exists in a 2d quantum bosons interacting with strong transverse
gauge fluctuations\cite{lee96,chen97}. 

In conclusion, we have calculated the change in magnetization and other
properties of the flux lattice in a model high-T$_c$ superconductor near and
above the melting transition {\em at relatively low fields}\cite{lowb}.  
The transition proceeds in two stages, 
a lower melting transition being followed by
a broader transition or crossover associated with 
proliferation of vortex loops in 3D, or vortex-antivortex 
pairs in 2D\cite{movie}. 
The magnetization increases sharply at this second
stage, in good agreement with recent experimental data. 

This work was supported by DOE Grant DE-FG02-90 ER45427
through the MISCON, and by NSF
Grant DMR94-02131.

\newpage

\newpage

\newpage

\newpage


\begin{references}
\bibitem{lala} We observe that $T_{\ell}$ is signaled by the appearance of an
``infinite vortex tangle'' and by large fluctuations in transverse vorticity,
not just by line entanglement and the occurrence of finite vortex loops, both 
setting in near $T_m$ for thermodynamic samples.
\bibitem{lowb} For highly anisotropic systems with $J_{xy}/J_z 
\sim (100-1000),$
we estimate $f=1/24$ roughly corresponds to $B\sim (250-25)$ 
Gauss\cite{hetzel92,koshelev97}. 
Furthermore, the density of unit vortex segments $|n_z|$ 
belonging to the field-induced vortices for 
$f=1/24$ for $T < T_m$ is less than 50 \% of those due to vortex loops 
induced at $T_{XY}$.
\end{references}
\end{document}